\documentclass[manuscript]{acmart}
\AtBeginDocument{%
  \providecommand\BibTeX{{%
    \normalfont B\kern-0.5em{\scshape i\kern-0.25em b}\kern-0.8em\TeX}}}

\setcopyright{acmlicensed}
\copyrightyear{2018}
\acmYear{2018}
\acmDOI{XXXXXXX.XXXXXXX}

\acmConference[AutoUI '24]{Make sure to enter the correct
  conference title from your rights confirmation emai}{June 03--05,
  2018}{Woodstock, NY}
\acmISBN{978-1-4503-XXXX-X/18/06}

\begin{document}

\title{A Survey of Driver Distraction and Inattention in Popular Commercial Software-Defined Vehicles}

\author{Lingyu Zhao}
\author{Yuankai He}

% \affiliation{%

% \affiliation{%

% \affiliation{%

% \affiliation{%

% \affiliation{%

% \affiliation{%

% \affiliation{%

\renewcommand{\shortauthors}{xxx and xxx, et al.}

\begin{abstract}
  As the automotive industry embraces software-defined vehicles (SDVs), the role of user interface (UI) design in ensuring driver safety has become increasingly significant. In crashes related to distracted driving, over 90\% did not involve cellphone use but were related to UI controls. However, many of the existing UI SDV implementations do not consider Drive Distraction and Inattention (DDI), which is reflected in many popular commercial vehicles. This paper investigates the impact of UI designs on driver distraction and inattention within the context of SDVs. Through a survey of popular commercial vehicles, we identify UI features that potentially increase cognitive load and evaluate design strategies to mitigate these risks. This survey highlights the need for UI designs that balance advanced software functionalities with driver-cognitive ergonomics. Findings aim to provide valuable guidance to researchers and OEMs to contribute to the field of automotive UI, contributing to the broader discussion on enhancing vehicular safety in the software-centric automotive era.
\end{abstract}

\begin{CCSXML}
<ccs2012>
   <concept>
       <concept_id>10003120.10003121.10003122</concept_id>
       <concept_desc>Human-centered computing~HCI design and evaluation methods</concept_desc>
       <concept_significance>500</concept_significance>
       </concept>
   <concept>
       <concept_id>10003120.10011738.10011776</concept_id>
       <concept_desc>Human-centered computing~Accessibility systems and tools</concept_desc>
       <concept_significance>300</concept_significance>
       </concept>
 </ccs2012>
\end{CCSXML}

\ccsdesc[500]{Human-centered computing~HCI design and evaluation methods}
\ccsdesc[300]{Human-centered computing~Accessibility systems and tools}


\keywords{AutomotiveUI, User Experience, User Interface, Driver Distraction and Inattention, Software-Defined-Vehicles, SDV}



\maketitle

\section{Introduction} \label{sec: introduction}

Driver distraction and inattention have been central to driver and vehicle safety in recent decades. While many states have passed and enforced laws relating to cellphone use, over 90\% of crashes related to distracted driving do not involve cellphone use \cite{RefWorks:RefID:7-2023traffic, RefWorks:RefID:3-distracted}.

Driver distraction and inattention are not new problems in road safety. There are numerous literature reviews on the dangers of distracted driving and driver inattention\cite{REGAN20111771, arun2012, sheila2005, RefWorks:RefID:8-stutts2003driver, kashevnik2021, strayer2007} and many countermeasures to mitigate driver inattention using sensors and machine learning models\cite{rienerautoui13, yun2014, katja2017, craye2015driver, wollmer2011}.

However, crashes relating to distracted driving have not been declining. The advent of software-defined vehicles (SDVs) has contributed little, if not exacerbated the problem. SDV heralds a new era in which software, rather than hardware, becomes the primary driver of vehicle functionality, innovation, and differentiation. This marks a significant transformation by decoupling software and hardware and enabling faster development and deployment. As commercial vehicles transition from traditional hardware-centric models to sophisticated platforms dominated by software, the \textit{user interface (UI)} assumes an increasingly central role in mediating the interaction between driver and vehicle. This shift, while fostering unprecedented levels of functionality, connectivity, and personalization, concurrently amplifies challenges associated with ensuring driver safety and minimizing distraction. 

\textit{\textbf{The integration of extensive software functionalities within the UI introduces complexities that can potentially overload the driver’s cognitive capabilities, leading to increased risks of distraction and inattention}}.

To the best of our knowledge, we did not find any publicly available resources that survey or discuss automotive UI's impact on driver distraction and inattention in the era of SDVs.

Therefore, to bridge this gap, this paper aims to provide a detailed survey of popular commercial vehicles, focusing on those that epitomize the software-defined paradigm. By evaluating the UI designs of these vehicles, this paper aims to identify specific aspects and features that contribute to driver distraction and inattention. Furthermore, this paper assesses the efficacy of various UI design strategies in mitigating such risks. Through this survey, we seek to delineate the implications of the shift towards software-centric automotive design on driver safety and propose potential pathways for aligning rapid technological evolutions with road safety.

In addressing these questions, we hope that this survey can shed light on the current state of automotive UI design amidst the rise of SDVs and contribute to the ongoing debate on balancing the benefits of software-driven functionalities with the need to safeguard drivers against distraction and inattention. By offering insights into how future UI designs can be optimized for safety without diminishing the user experience, this paper endeavors to chart a course toward a new generation of vehicles that balances innovation with driver well-being.

The structure of the paper is as follows: Section \ref{sec: bg} provides important background information on Software-Defined-Vehicles (SDV) and jarring statistics on driver distraction and inattention. Section \ref{sec: motivation} outlines the motivation for our survey of commercially available passenger vehicles. Section \ref{sec: setup} details how the survey is conducted and what aspects of the vehicle are placed under scrutiny. Next, Section \ref{sec: results} demonstrates the survey results, and outlines areas of improvement and areas of good design. Finally, Section \ref{sec: conclusion} concludes the paper.
\section{Background and Related Works} \label{sec: bg}
This section highlights the danger driver distraction and inattention poses and offers some background knowledge on how software-defined-vehicles (SDVs) may contribute or mitigate such danger.

\subsection{Crash Data on Driver Distraction and Inattention}

\begin{table}[!ht]
  \caption{Traffic Crashes and Distraction-Affected (D-A) Crashes}
  \label{tab:crashstats}
  \begin{tabular}{l|c|c|c}
    \toprule
    Year & Total & D-A Crashes & \% of D-A Crashes not Related to Cellphone Use \\
    \midrule
    2019 & 6,756,084 & 986,204 & 94 \\
    2020 & 5,251,006 & 680,305 & 91 \\
    2021 & 6,102,936 & 804,928 & 92 \\
    \bottomrule
  \end{tabular}
\end{table}

Shown in Table \ref{tab:crashstats}, according to NHTSA's report in May 2023, even with multiple states banning cell phone use while driving and other activities that distract the driver, distracted-affected crashes accounted for 15\% of total crashes in 2019, 13\% in 2020, and 13\% in 2021, resulting in 986,204, 680,305, and 804,928 crashes respectively \cite{RefWorks:RefID:7-2023traffic, RefWorks:RefID:3-distracted}. \textbf{\textit{It is important to note that over 90\% of D-A crashes are not related to cellphone use but are related to the driver eating, drinking, or operating vehicle UI controls.}}

\subsection{Software-Defined-Vehicles (SDVs)}
SDV is the prodigal child of current OEMs. SDV promises the vision of decoupling software and hardware dependencies and enables faster development, testing, and deployment. It also promises the end-users continuing updates to the vehicle in the years following their purchase.

SDV has created a shift in a vehicle's functional hardware, electrical/electronic architecture (EEA), computing platform, operating system (OS) kernel, middleware, service layer of service-oriented architecture (SOA), functional application, service application, and cloud service platform \cite{RefWorks:RefID:4-liu2022impact}. This shift creates unprecedented levels of functionality in the instrument cluster and personalization in the vehicle's infotainment center, \textbf{\textit{resulting in a situation where the push for newer software functionalities often overshadows the concern for driver distraction and inattention.}}

\section{Motivation} \label{sec: motivation}

 Despite the rapid advancement and integration of SDVs into the automotive market, there remains a lack of comprehensive studies that dive into how these vehicles' user interface (UI) designs contribute to or mitigate driver distraction and inattention. This oversight is significant, as the shift towards SDVs promises to redefine vehicular interaction through software rather than hardware. Existing design often overlooks the critical aspect of driver safety. This survey seeks to bridge this knowledge gap by systematically analyzing the UI designs of popular commercial passenger vehicles and assessing their impact on driver focus and safety. This survey aims to underscore the necessity for UI designs that prioritize functionality and safety.
\section{Experiment Setup} \label{sec: setup}
The primary objective of this study is to evaluate the influence of various in-vehicle controls on Driver Distraction and Inattention (DDI). Recognizing the critical role that user interface controls play in overall driving safety, the research focused on those controls most frequently interacted with by drivers during vehicle operation. Specifically, the study examines the design and functionality of controls on the central console and instrument cluster, steering wheel, and transmission system. These components were chosen due to their high likelihood of driver interaction while driving, thus representing significant areas of interest for assessing potential DDI risks.

\subsection{Role and Perspective}

\begin{figure}[!ht]
  \centering
  \includegraphics[width=0.8\textwidth]{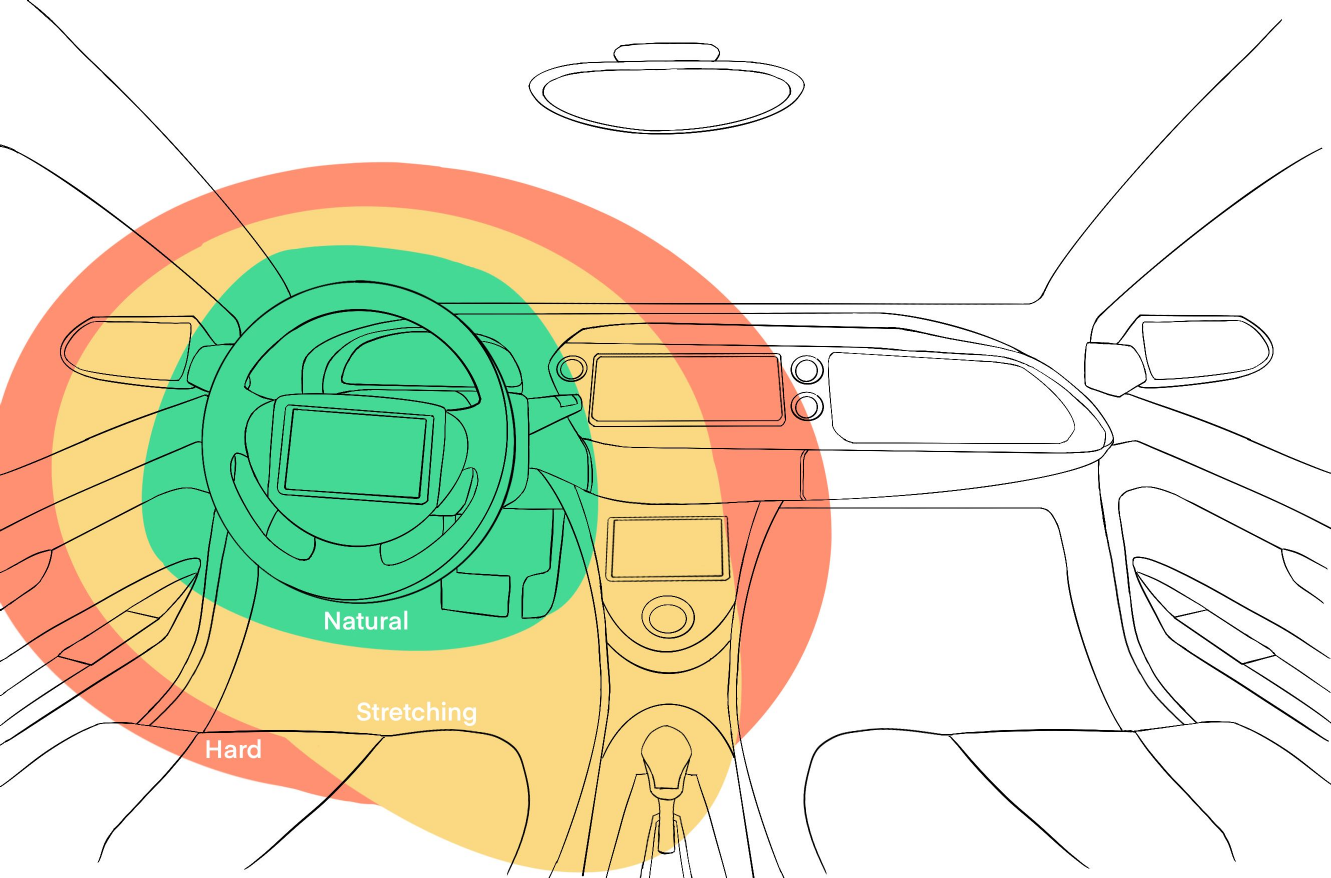}
  \caption{Three Zones for Driver Interaction Difficulties}
  \label{fig:armzone}
\end{figure}

The survey is conducted from a driver's perspective to authentically assess the ergonomic and cognitive demands placed on individuals while operating the vehicle. This approach ensured the evaluation reflected realistic usage scenarios and drivers' interactions with the vehicle's controls under typical driving conditions.

As shown in Fig \ref{fig:armzone}, the cockpit is divided into three distinct regions based on the interaction difficulty. The area closest to the driver, \textit{Natural}, is marked in green; the area further away but still easy to interact, \textit{Stretching}, is marked in yellow; and the area furthest away to interact, \textit{Hard}, is marked in red.

\subsection{Areas of Focus}

\begin{enumerate}
    \item \textbf{Steering Wheel and Instrument Cluster:} The investigation into the steering wheel controls and the associated instrument cluster emphasizes the information architecture of these controls. Key points of interest included the potential for accidental activation of controls, mechanisms available for easily canceling unintended inputs, the likelihood of misunderstanding the functions of specific controls, and the overall accessibility of desired controls while driving. The goal was to identify design elements that either contribute to or mitigate driver distraction and inattention.
    \item \textbf{Central Console:} The central console analysis concentrates on the methods available for interacting with its controls. Of particular interest were the various modalities provided for engaging with a single control to fulfill different functions. This examination aimed to understand how multipurpose controls and their interaction methods could influence driver focus, considering both the benefits of such design choices and the potential for causing confusion or distraction.
    \item \textbf{Transmission:} With respect to the vehicle's transmission controls, the focus was on the significance of maintaining physical controls for critical functions such as gear selection. Additionally, the clarity and prominence of gear status indications were evaluated to determine their effectiveness in conveying essential information to the driver without diverting attention away from the driving task. This analysis sought to underscore the importance of intuitive physical controls and clear feedback mechanisms in minimizing DDI.
\end{enumerate}

\subsection{Evaluation Criteria}

Each of the aforementioned vehicle control areas is assessed based on a set of criteria designed to evaluate their impact on driver distraction and inattention. These criteria included ergonomic design, cognitive load, intuitiveness, error forgiveness, and feedback quality. The study utilized a combination of direct observation, tactile feedback, and cognitive assessment to examine how each control interface aligns with best practices for minimizing DDI. Through this comprehensive approach, the research aims to highlight design strengths and pinpoint areas where improvements could significantly enhance driver safety and interface usability.

\section{Survey Result} \label{sec: results}
The survey covers 13 of the best-selling automotive brands in the US market, which account for 83\% of all new car sales from January to March 2024 in the United States\cite{RefWorks:RefID:9-usa}.
We surveyed 17 models across these 13 best-selling automotive brands, revealing insightful results regarding the design of user interface controls and their impact on Driver Distraction and Inattention (DDI). This section highlights key findings from the evaluation, categorized by the focus areas: steering wheel and instrument cluster, central console, and transmission. Specific examples from the surveyed models are noted, with actual brands and models removed.

\subsection{Good Design Practices Mitigating DDI}
This subsection lists some of the UI/UX designs that were exceptional or were very smart about driver distraction and inattention when surveying commercially available vehicles.

\textbf{Steering Wheel:}

\begin{figure}[!ht]
\centering
\begin{minipage}[t]{0.45\textwidth}
  \centering
  \includegraphics[width=0.8\linewidth, angle=270]{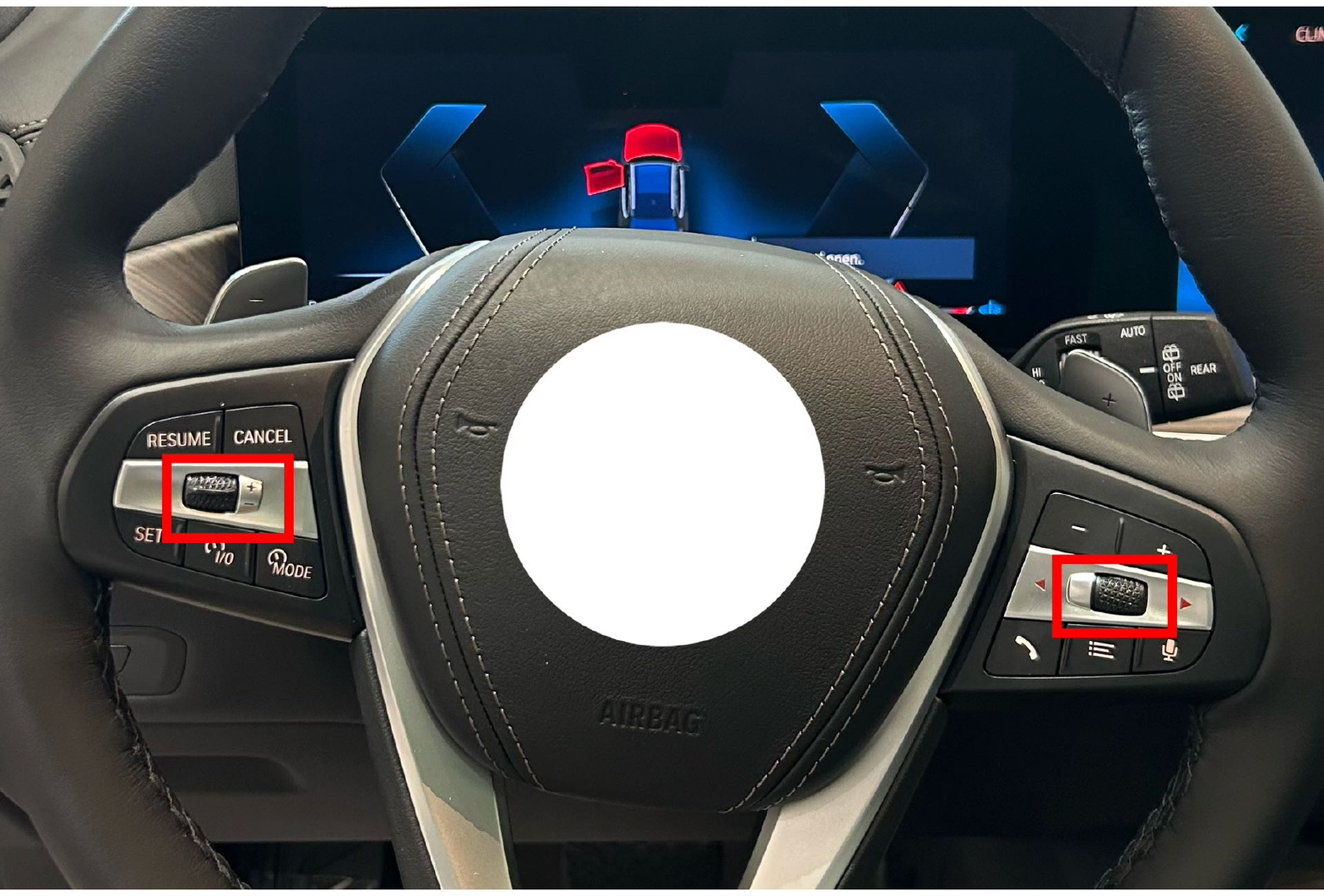}
  \caption{Multifunctional buttons on a steering wheel, highlighting scrolling and pressing functions with tactile cues for operation.}
  \label{fig:button}
\end{minipage}%
\hspace{0.04\textwidth} 
\begin{minipage}[t]{0.45\textwidth} % Use the [t] option to align the top of each minipage
  \centering
  \includegraphics[width=0.8\linewidth, angle=270]{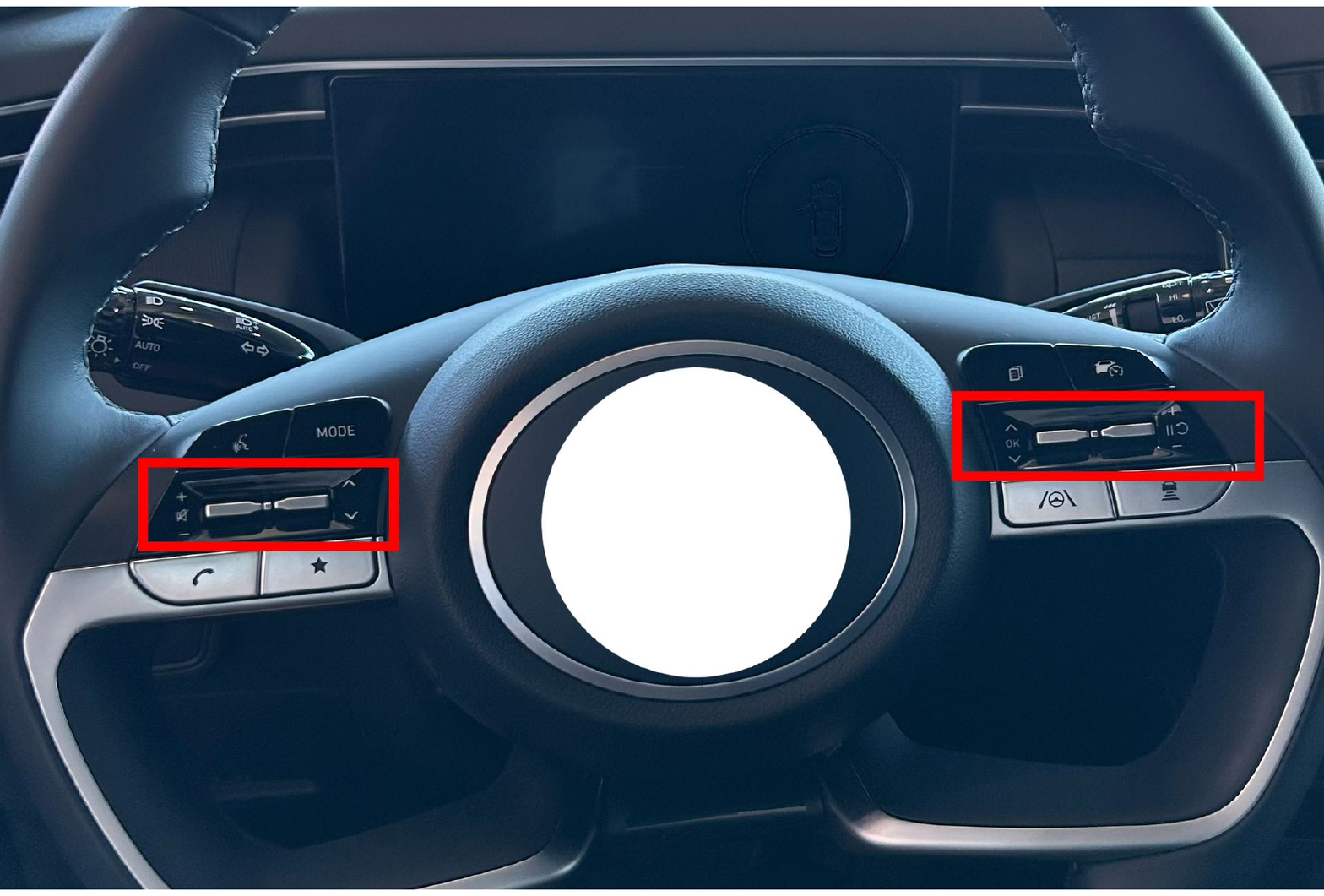}
  \caption{The rocker switch is a compact and functional alternative to conventional knobs.}
  \label{fig:rocker}
\end{minipage}
\end{figure}

\begin{figure}[!ht]
  \centering
  \includegraphics[width=0.4\textwidth, angle=270]{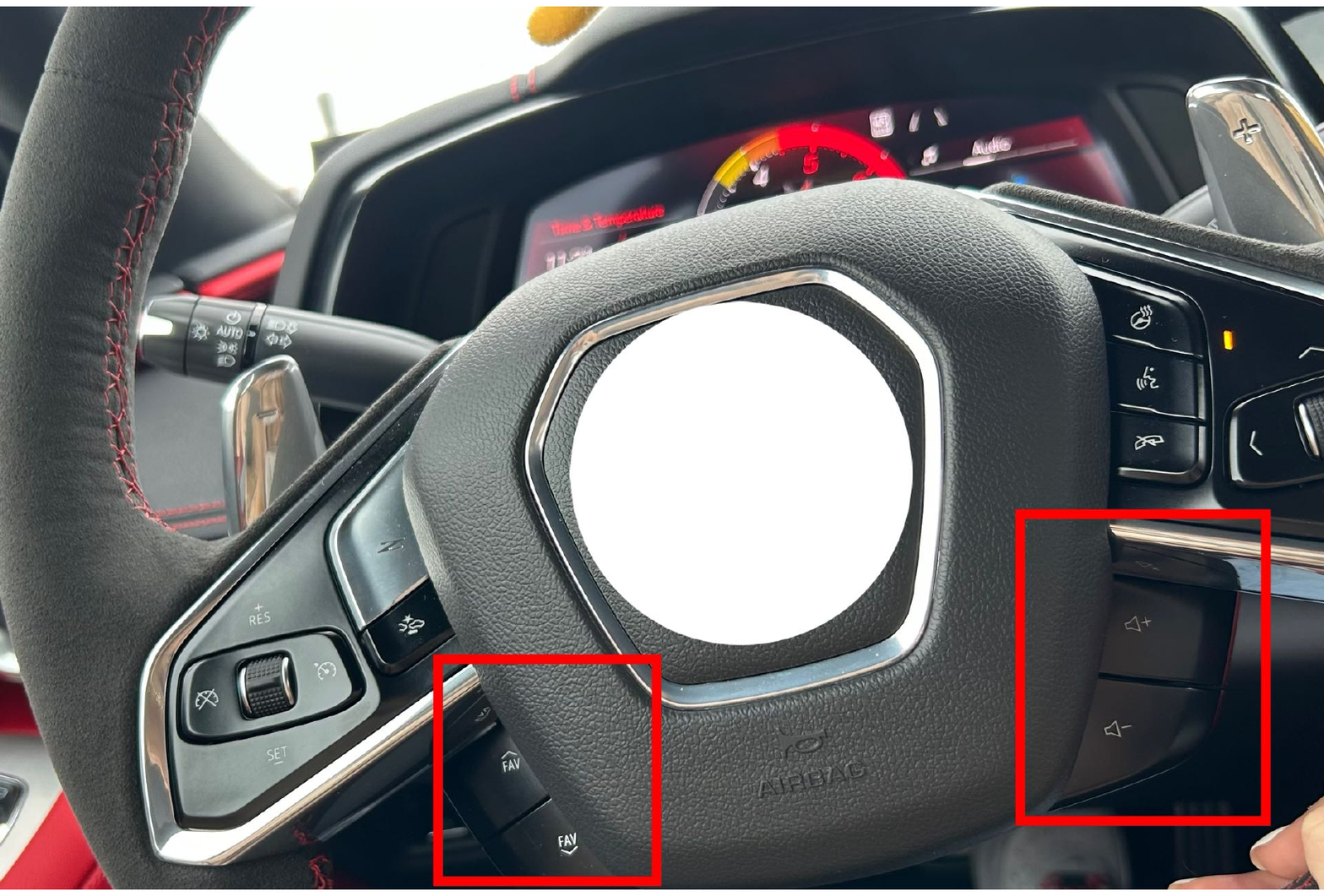}
  \caption{Erogonomic, distraction-free button design.}
  \label{fig:buttonmiddlefingeract}
\end{figure}

Integrating multifunctional buttons into vehicle steering wheel designs is significant in automotive ergonomics and functionality. Numerous manufacturers have adopted this strategy to enable drivers to perform multiple actions with a single control element, such as scrolling upwards or downwards and pressing down to activate distinct functions. For example, as illustrated in Figure \ref{fig:button}, the button on the steering wheel's right-hand side allows for volume adjustment through scrolling actions, while a press toggles the mute function on or off. A similar configuration is found on the left side, controlling cruise control speed; however, it uniquely features a slender bar on top, indicating that it supports scrolling but not pressing actions.

This design cleverly enhances usability while minimizing driver distraction. Drivers can operate these buttons intuitively, with little need to divert their gaze from the road, as each interaction is logically linked to its function and straightforward to execute. Additionally, this design approach conserves space on the steering wheel, ensuring sufficient separation between buttons to prevent accidental presses. The presence of a slender tactile bar not only serves as a functional guide on how to use the button but also increases friction, further reducing the likelihood of unintended interactions. Should a mistouch occur, drivers can easily click on the button again to cancel the action, adding an extra layer of user forgiveness and functionality. The tactile feedback and strategic placement of these controls contribute to a safer driving experience by allowing users to maintain focus on the road while effortlessly managing vehicle functions. This innovation underscores the automotive industry's commitment to enhancing user experience and safety through thoughtful design.

An alternative to traditional knobs, the rocker switch, shown in Fig \ref{fig:rocker}, is highlighted for its space efficiency and functionality, serving as an effective design solution in multiple brands. This subtle design element intuitively communicates the button's functionality, exemplifying a thoughtful integration of form and function to guide user interaction. Each switch is clearly labeled with its own icon, allowing users to easily understand how to use it. Although it allows the driver to access a broad range of information, enhancing utility, the proximity of the rocker switches to each other can occasionally lead to mistouches. This design challenge highlights the delicate balance required between space-saving and usability, underscoring the importance of ergonomic design to minimize potential errors while maximizing functionality. Since this model has more controls on the steering wheel than average, it may take some time for drivers to familiarize themselves. Because they are all physical buttons, the force and haptic feedback provided is very clear, aiding in quick learning and error recovering.

A notable design feature, shown in Fig \ref{fig:buttonmiddlefingeract}, is observed where additional buttons are ingeniously positioned to be activated by an upward press from either the index or middle finger during driving. This arrangement carefully spaces each button to minimize the risk of accidental activation while simultaneously expanding the driver's ability to access a broader range of functions. These are physical controls in larger sizes, providing even more force feedback and larger icons than regular steering wheel buttons, which enhances tactile interaction and ease of use. Crucially, this design allows for such interactions without requiring the driver to divert their gaze from the road, enhancing operational safety and convenience. If a mistouch occurs, drivers can press the button next to it to redo the interaction, making correcting unintended inputs straightforward. Using the index or middle finger to interact with these controls also requires the driver to maintain a tight grip on the steering wheel, which is a constant reminder to keep good control of the vehicle and stay attentive to road conditions. This integration of function and safety design subtly reinforces responsible driving habits.

\textbf{Central Console:}

\begin{figure}[!ht]
  \centering
  % First image minipage taking up 60% of text width, centered vertically
  \begin{minipage}[c]{0.60\textwidth}
    \centering
    \includegraphics[width=0.83\textwidth, angle=270]{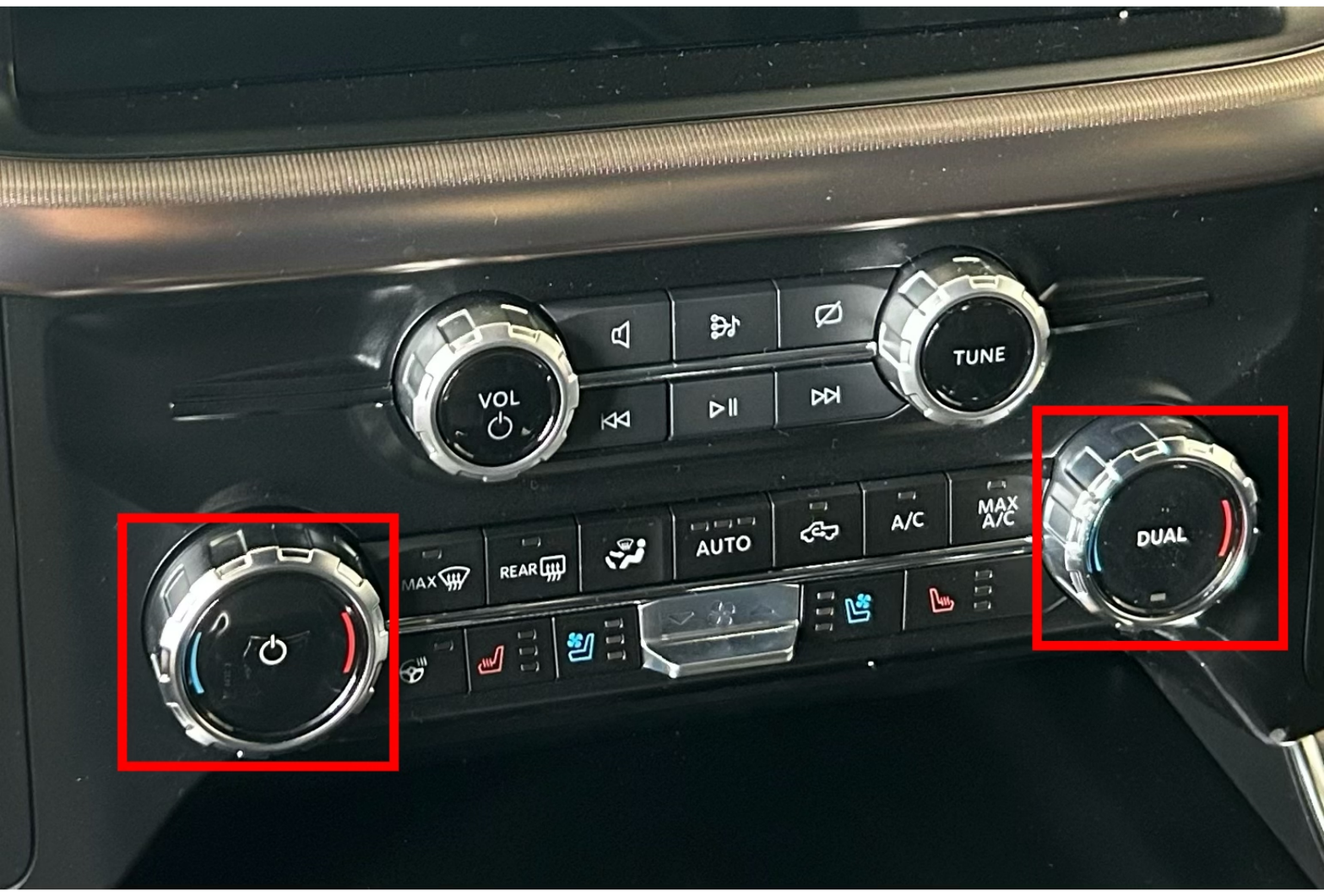} % Adjust width as needed
    \caption{Multi-functional knobs with intuitive color coding and iconography for simplified use.}
    \label{fig:colorcoded}
  \end{minipage}
  \hfill % This will add space between the two minipages
  % Second image minipage taking up 30% of text width, centered vertically
  \begin{minipage}[c]{0.32\textwidth}
    \centering
    \includegraphics[width=\textwidth, keepaspectratio]{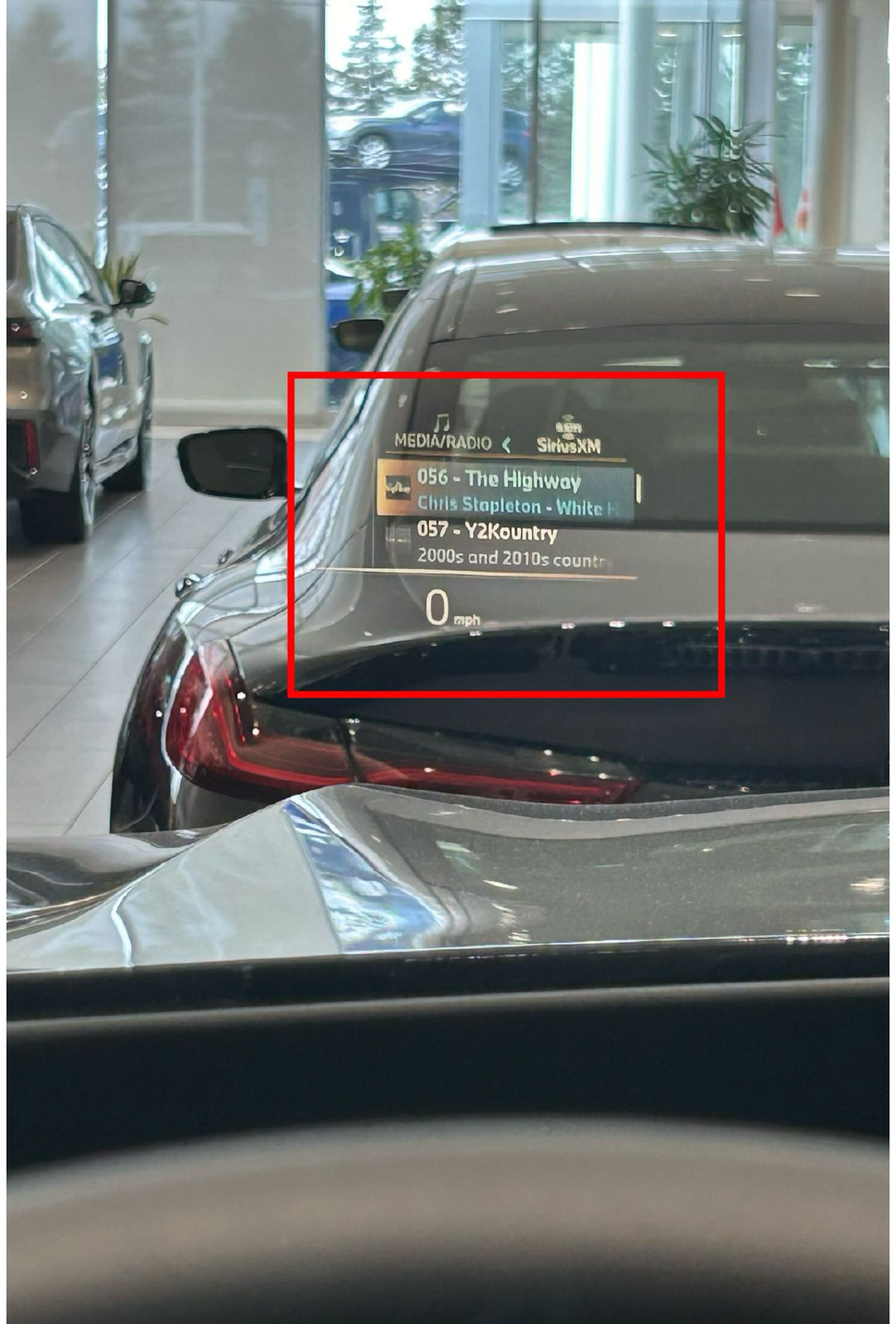}
    \caption{A heads-up display serves as an extended instrument cluster for customizable information display.}
    \label{fig:hud_display}
  \end{minipage}
\end{figure}

The use of knobs for adjusting various settings such as volume, temperature, and wind speed has been a common feature that received positive feedback for enhancing the user experience (UX) in vehicle interfaces. These controls employ multiple interaction methods — rotational (clockwise and counterclockwise) and pressing — providing users precise control over adjustments while minimizing the cognitive load associated with operating complex systems. Compared to switches, knobs allow drivers to adjust climate or volume in a wider range within a shorter time with a single twist, effectively decreasing the chance of Driver Distraction and Inattention (DDI). They also offer clear feedback on the extent of adjustments based on the force and haptic feedback from the knob, enhancing usability. The inclusion of clear visual cues, such as color coding and iconography surrounding the knobs, plays a critical role in their intuitiveness. For instance, as illustrated in Fig \ref{fig:colorcoded}, the specific design of climate control knobs utilizes these elements to convey functionality immediately, allowing users to adjust settings almost instinctively without the need to divert their focus from the road. Additionally, with the color-coded indication, drivers can easily understand how to use it and also recover from any errors quickly. Its location very close to the steering wheel enables the driver to quickly take off their right hand, adjust climate settings, and put it back on the steering wheel, further enhancing safety and convenience.

Expanding on intuitive design, Fig \ref{fig:hud_display} highlights an innovative heads-up display (HUD) approach. This feature acts as an extension of instrument cluster, presenting customizable information directly within the driver’s line of sight. By projecting crucial data such as speed, navigation prompts, and safety warnings onto the windshield, the HUD significantly reduces Driver Distraction and Inattention (DDI). This design ensures that important information is more visible and allows drivers to maintain their attention on the driving environment, thereby enhancing overall vehicle safety. 

% The seamless integration of the HUD with traditional controls like knobs further exemplifies a user-centered design philosophy that prioritizes accessibility and ease of use, reflecting a sophisticated understanding of human factors in automotive UX design.

\textbf{Transmission:}

\begin{figure}[!ht]
  \centering
  \includegraphics[width=0.2\linewidth]{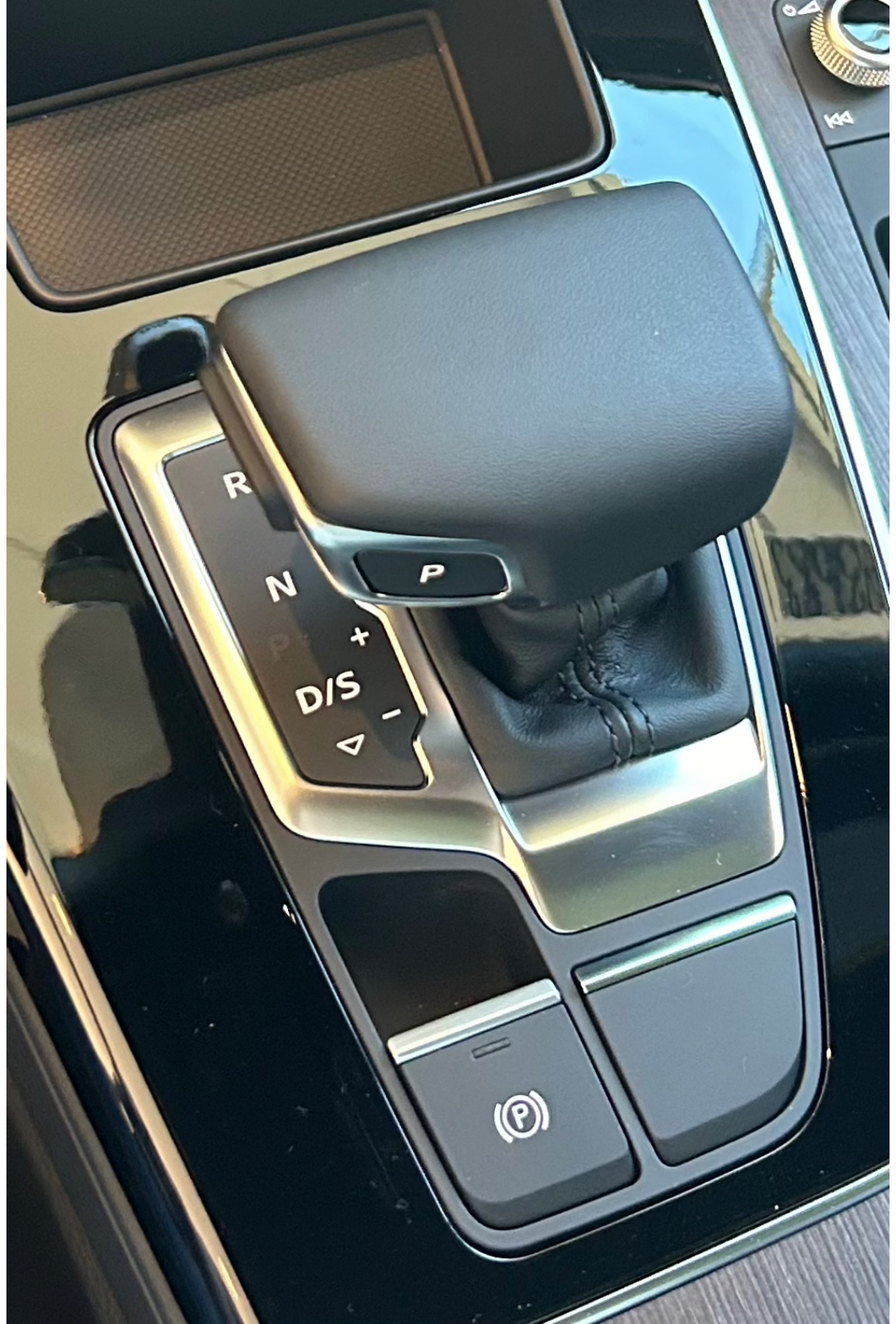} % Use full width of the minipage
  \caption{Innovative transmission control with a distinct 'Park' button and intuitive 'Drive' and 'Reverse' actions.}
  \label{fig:transmission_control}
\end{figure}

Innovative transmission control designs were noted for enhancing user interaction by simplifying operations and minimizing cognitive effort. One design, as featured in Fig \ref{fig:transmission_control}, incorporates a distinct button for 'Park' atop a rectangular casing, complemented by intuitive actions for 'Drive' and 'Reverse'. This ergonomic design significantly eases the gear selection process for the driver by separating 'Park' from 'Drive' and 'Reverse', thus reducing the cognitive load. Drivers no longer need to worry about how much force to apply or repeatedly check their selection, allowing more focus on driving. The intuitive nature of the control layout means that the functions are easy to understand and use, minimizing the likelihood of errors. The feedback quality from these controls is also meticulously designed; tactile feedback confirms engagement with a satisfying click, and visual indicators on the control panel clearly show the current gear mode. This multi-modal feedback system reinforces the driver’s actions, ensuring correct operation without the need to divert their gaze.

\subsection{Areas for Improvement}
This subsection discusses and identifies improvements OEMs should consider when designing automotive UI focused on mitigating DDI.

\begin{figure}[!ht]
  \centering
  % First image minipage taking up 60% of text width, centered vertically
  \begin{minipage}[c]{0.60\textwidth}
    \centering
    \includegraphics[width=0.83\textwidth, angle=270]{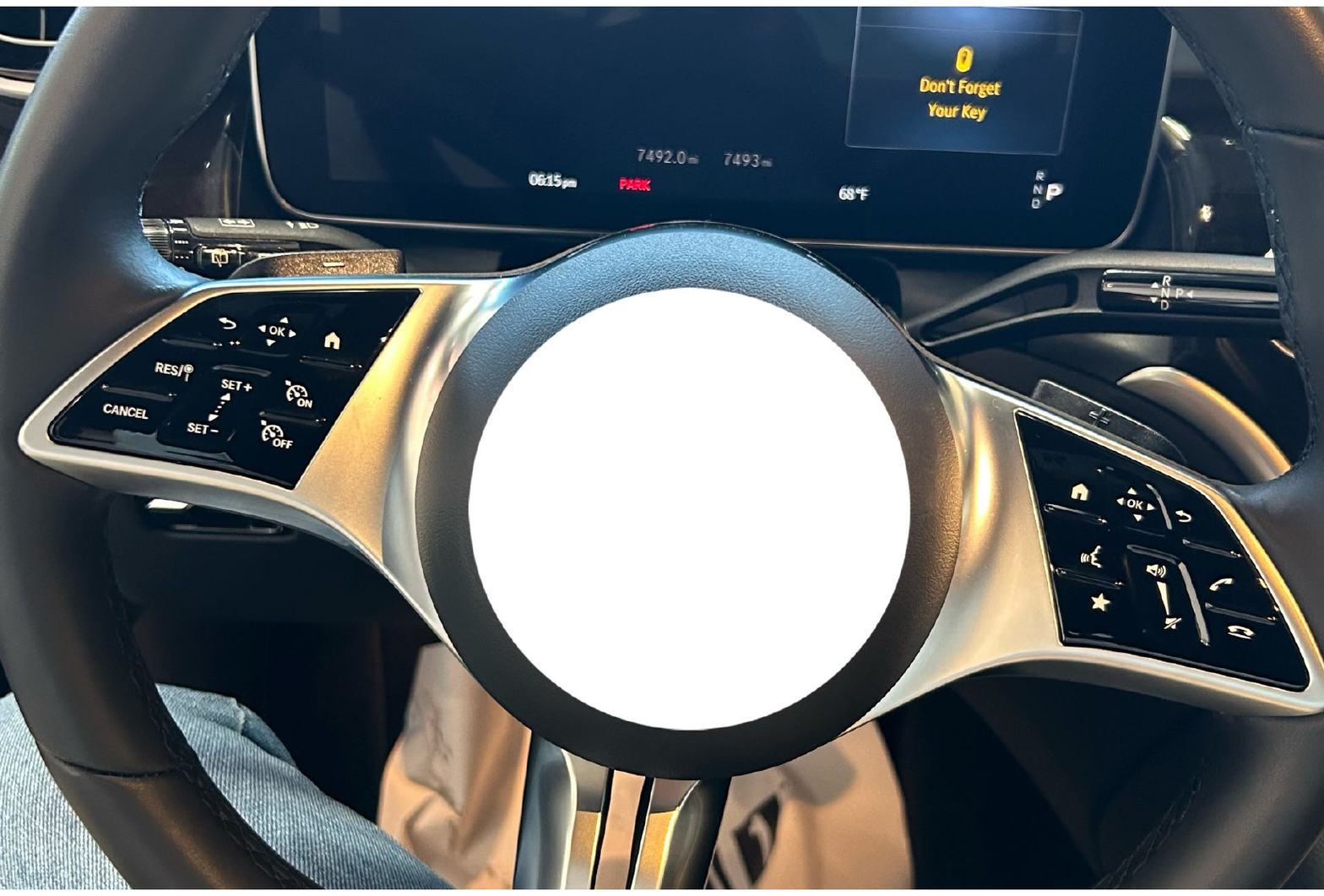} % Adjust width as needed
    \caption{Steering wheel digital controls highlighting the challenge of accidental activation and the need for improved sensitivity and feedback mechanisms.}
    \label{fig:steering_wheel_controls}
  \end{minipage}
  \hfill % This will add space between the two minipages
  % Second image minipage taking up 30% of text width, centered vertically
  \begin{minipage}[c]{0.32\textwidth}
    \centering
    \includegraphics[width=\textwidth, keepaspectratio]{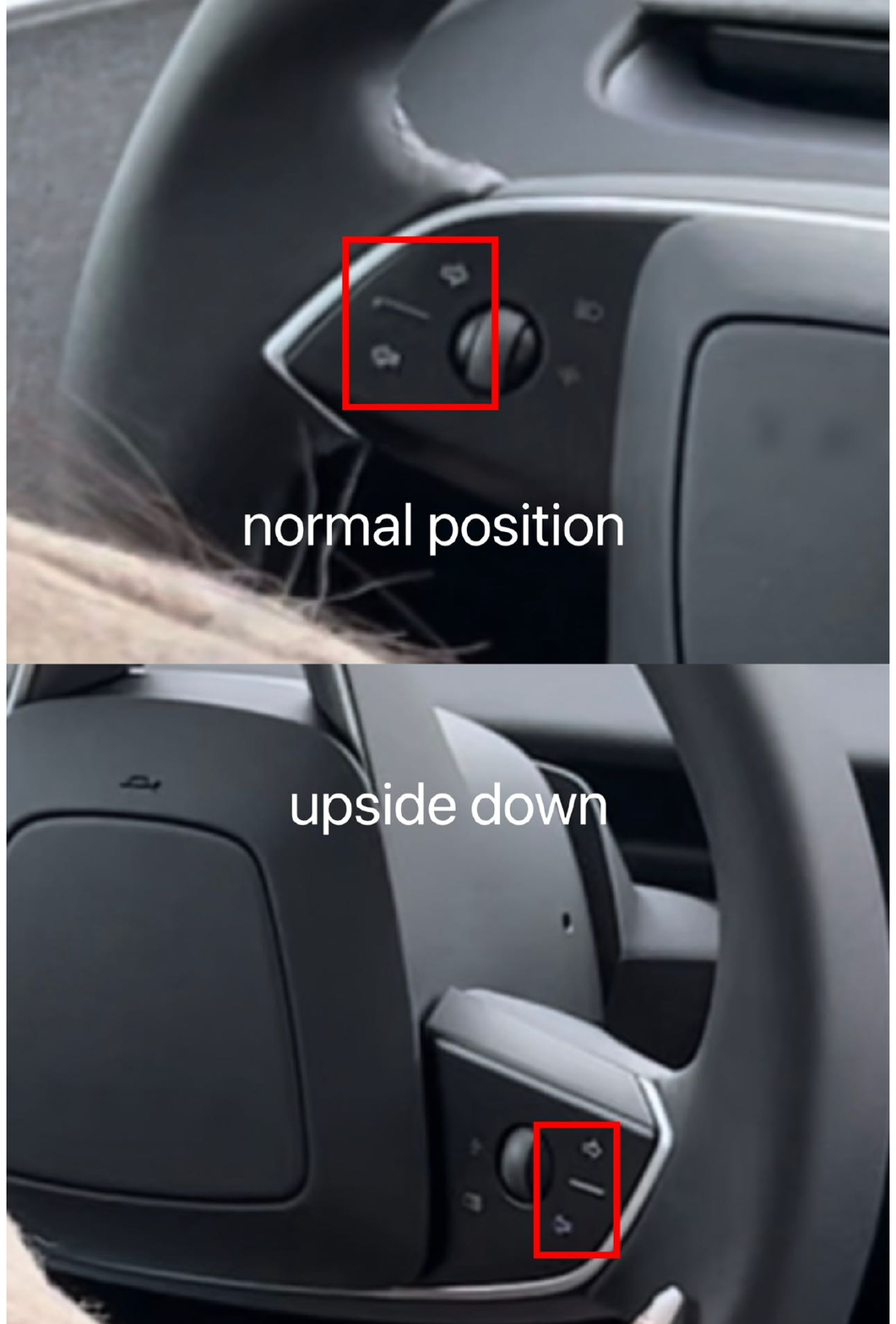}
    \caption{Steering wheel digital controls highlighting the challenge of accidental activation and the need for improved sensitivity and feedback mechanisms. Top: Steering Wheel at 0$^{\circ}$; Bottom: Steering Wheel at 180$^{\circ}$}
    \label{fig:steering_wheel_turn_signal}
  \end{minipage}
\end{figure}

\textbf{Steering Wheel:}

Implementing digital controls on the steering wheel, as shown in Fig \ref{fig:steering_wheel_controls}, introduces a potential distraction due to the difficulty in distinguishing these inputs from road-induced vibrations, despite having haptic feedback. This setup, featuring both pressing and sliding actions, increases cognitive load as drivers must remember how each button functions, compromising the intuitiveness of the interface. The presence of a "go back" button offers some error forgiveness, allowing for easy correction of mistakes, but the feedback quality still needs enhancement to ensure that haptic signals are distinctly perceivable from unrelated vibrations. 

Challenges arise with the placement and labeling of controls, particularly turn signals on the steering wheel, which could become less intuitive when the wheel is not in its default position, shown in Fig \ref{fig:steering_wheel_turn_signal}, suggesting the need for designs that maintain consistent operation regardless of wheel orientation. For example, an inexperienced driver may have a hard time differentiating the left and right turn signal buttons when the steering wheel is at 0$^{\circ}$ and at 180$^{\circ}$. Such a scenario underscores the importance of ergonomic design that accommodates changes in the wheel's position, ensuring that critical controls are always accessible and clearly identifiable, thereby enhancing safety and reducing the cognitive load on drivers.

\textbf{Central Console:}

\begin{figure}[!ht]
  \centering
  % First image minipage taking up 60% of text width, centered vertically
  \begin{minipage}[c]{0.60\textwidth}
    \centering
    \includegraphics[width=0.83\textwidth, angle=270]{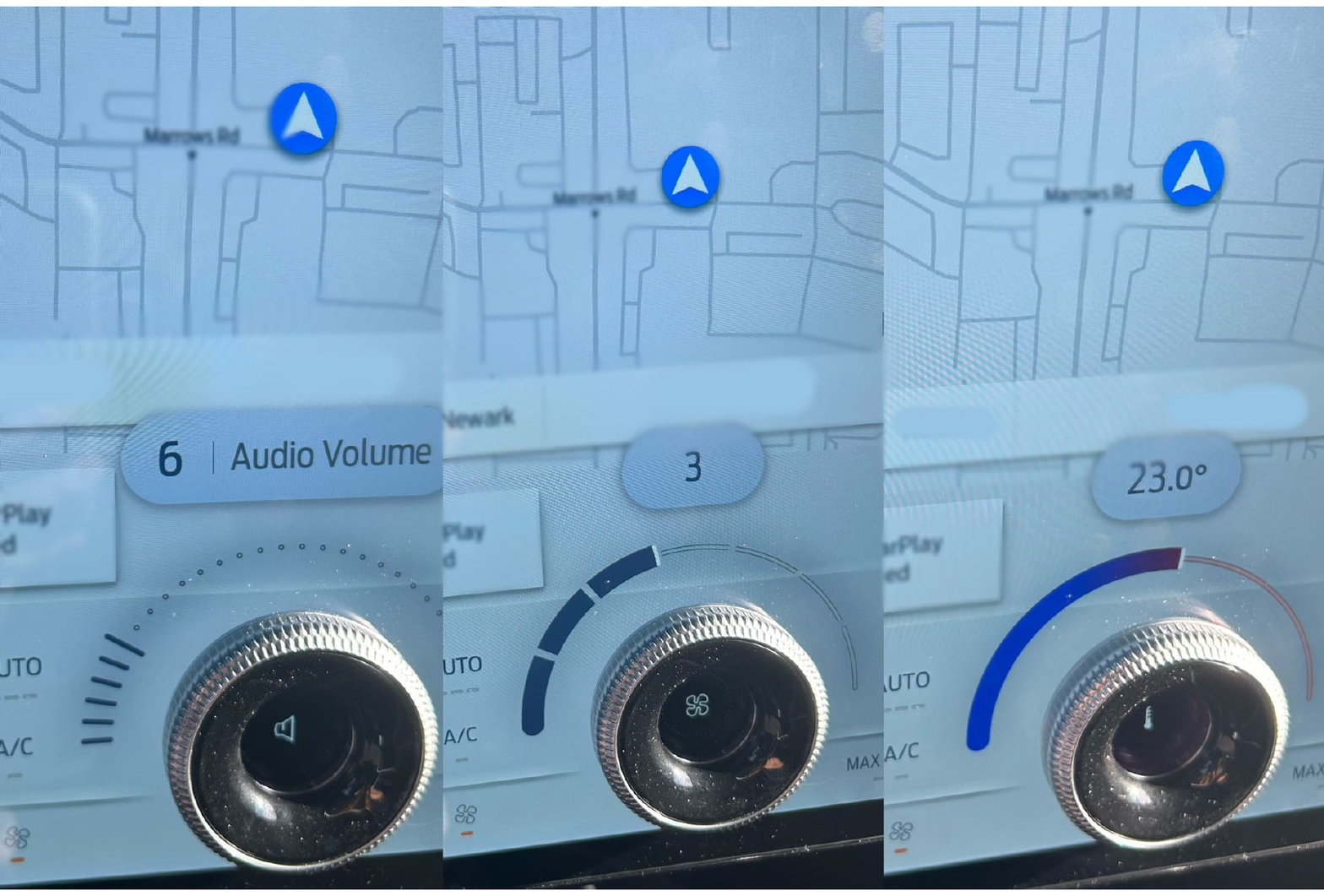} % Adjust width as needed
    \caption{Knobs designed for multi-functional adjustments.}
    \label{fig:multifunction_knobs}
  \end{minipage}
  \hfill % This will add space between the two minipages
  % Second image minipage taking up 30% of text width, centered vertically
  \begin{minipage}[c]{0.32\textwidth}
    \centering
    \includegraphics[width=\textwidth, keepaspectratio]{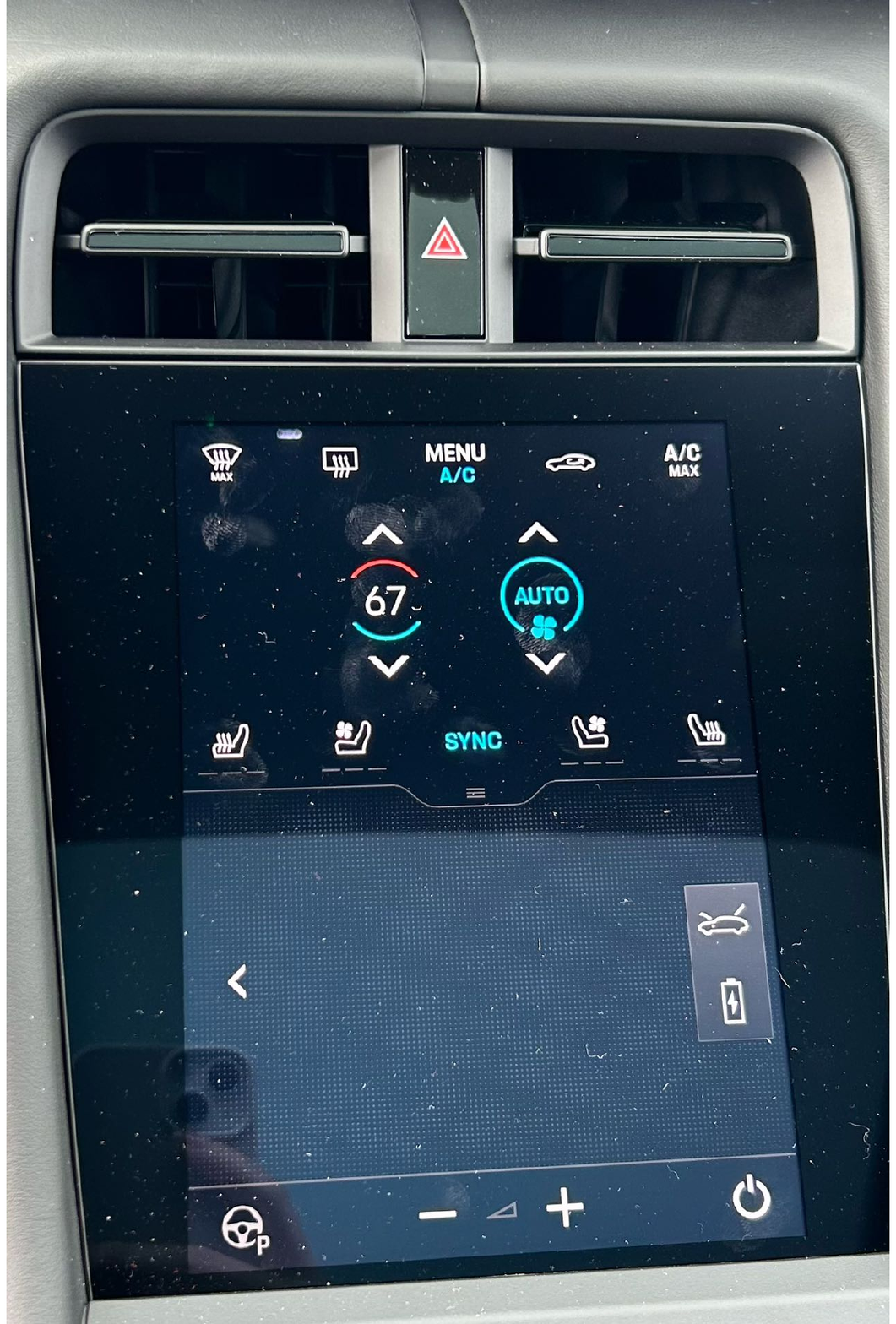}
    \caption{Screen-based climate control}
    \label{fig:climate_touchpad}
  \end{minipage}
\end{figure}

% Challenges arise with the placement and labeling of controls, particularly turn signals on the steering wheel, which can become less intuitive when the wheel is not in its default position, as shown in Fig \ref{fig:steering_wheel_turn_signal}. This issue highlights the need for designs that maintain consistent operation regardless of wheel orientation. For instance, an inexperienced driver may struggle to differentiate between the left and right turn signal buttons when the steering wheel is rotated from 0$^{\circ}$ to 180$^{\circ}$. 

The interface design, shown in Fig \ref{fig:multifunction_knobs}, features a physical knob attached to a digital screen, cleverly incorporates tactile feedback, yet presents a notable usability issue. The current selected option is indicated by an icon positioned within the hole of the knob, making it difficult for drivers to see from their driving position. This placement can significantly impair visibility and accessibility, forcing drivers to adjust their posture or take their eyes off the road to confirm their selection. Such a design choice not only diverts attention from driving but also increases cognitive load as drivers must focus more on deciphering the control settings rather than maintaining their focus on the road. Improving the visibility of these indicators, perhaps by relocating them to a more visible part of the screen or enhancing the contrast and size of the icons, could mitigate this issue and enhance the overall safety and ergonomics of the interface.

As illustrated in Fig \ref{fig:climate_touchpad}, screen-based controls for sequential option selection lead to increased interaction times and cognitive load. These designs often involve multiple steps for adjusting settings, which can become particularly cumbersome under bright lighting conditions that impair screen visibility. The ergonomic design requires drivers to divert their attention from the road to the screen, increasing cognitive burden. The interface's intuitiveness is limited, particularly for those unaccustomed to digital controls, and the lack of tactile feedback diminishes the quality of user feedback. Moreover, these systems offer poor error forgiveness; correcting a misselection requires additional navigation steps, further diverting attention from driving tasks. 

\textbf{Transmission:}

\begin{figure}[!ht]
  \centering
  \includegraphics[width=0.3\linewidth]{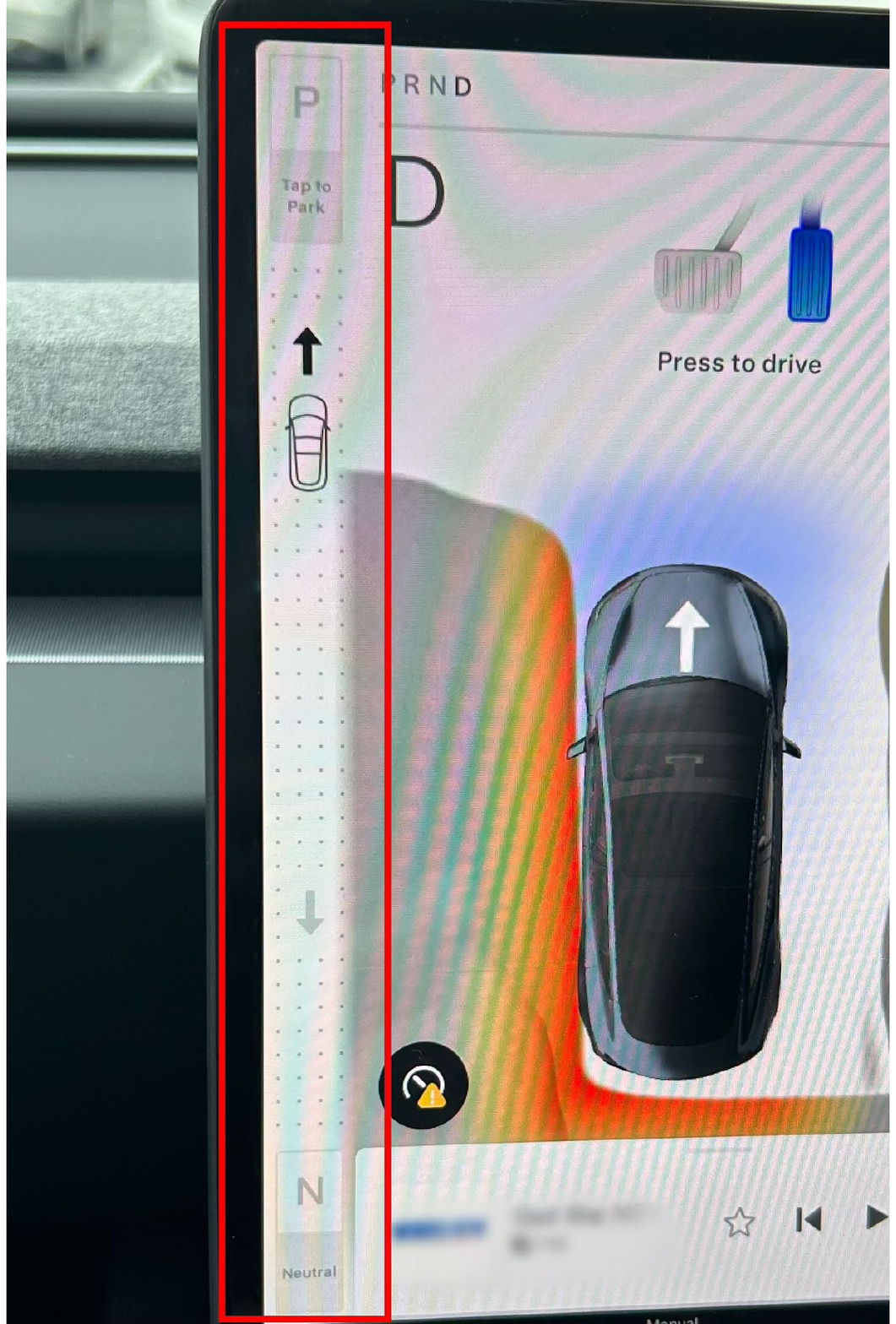} % Image path needs to be corrected as per your image location
  \caption{Screen-based transmission control.}
  \label{fig:screen_transmission}
\end{figure}

Adopting digital controls for critical functions like gear selection introduces significant risks, particularly in terms of system accessibility during malfunctions, as depicted in Fig \ref{fig:screen_transmission}. If the digital system fails or becomes unresponsive, drivers could find themselves unable to change gears, which can compromise safety. This issue highlights a major concern in ergonomic design—ensuring that controls are always accessible and functional, especially for critical operations.

Furthermore, these digital controls typically utilize gestures or touch-based inputs that might not be intuitive for all users, especially those familiar with traditional, physical gear levers. This can create confusion and operational delays, increasing the cognitive load as drivers must process how to use unfamiliar controls in potentially high-pressure situations. Such a setup lacks the immediacy and ease of use that physical controls offer, where the action and outcome are directly connected.

The touchscreen interface for gear selection shown in Fig \ref{fig:screen_transmission}, while aesthetically modern, lacks the tactile feedback that drivers rely on to operate controls without looking. This absence reduces the feedback quality, as there is no physical feeback confirming the selection. Drivers must instead rely on visual confirmation, which can divert their attention from the road, thus increasing the risk of errors and accidents.

Regarding error forgiveness, digital interfaces often provide limited immediate corrective options. Physical controls typically allow for quick, on-the-fly adjustments without looking, but digital controls may require navigating through menus or screens to correct an input, complicating rapid response and recovery from errors.

\section{Conclusion} \label{sec: conclusion}
With the automotive industry's shift towards SDVs, UI is becoming increasingly central in managing the interaction between the driver and the vehicle. This transition has led to both enhanced functionality and greater risks of distraction due to the complexity of UI designs. This work focuses on popular commercial vehicles, analyzing various UI elements — such as controls on the steering wheel, central console, and transmission — to determine their contribution to driver distraction and inattention. Based on this study, current challenges, trends, and future directions are listed. These insights aim to provide valuable guidance to researchers and OEMs to actively contribute to the field of automotive UI. The emphasis lies in balancing how future UI designs can be optimized for safety without diminishing the user experience

\section{Citations and Bibliographies}

% The use of \BibTeX\ for the preparation and formatting of one's
% references is strongly recommended. Authors' names should be complete
% --- use full first names (``Donald E. Knuth'') not initials
% (``D. E. Knuth'') --- and the salient identifying features of a
% reference should be included: title, year, volume, number, pages,
% article DOI, etc.

% The bibliography is included in your source document with these two
% commands, placed just before the \verb|\end{document}| command:
% \begin{verbatim}
\bibliographystyle{ACM-Reference-Format}
\bibliography{main}

\end{document}